\documentstyle[12pt,thmsa,sw20lart]{article}

\input{tcilatex}
\begin{document}

\author{{\Large V. A. Grigoryan\bigskip } \\
{\it Yerevan State University , Department of Physics , }\\
{\it 1 Alec Manoogian St., 375049, Yerevan, Armenia\bigskip }}
\title{{\LARGE The Influence of Structural Inhomogeneities on Intragranular
Properties of Y- and Bi-based Superconductors\bigskip }}
\date{Preprint YerPhi \# \bigskip 1518(18)-98}
\maketitle

\begin{abstract}
The measurements of temperature dependence of AC magnetic susceptibility
were used to study the inf{}luence of inhomogeneities on intragranular
superconducting properties of $YBa_{2}Cu_{3-X}M_{X}O_{Y}$
(M=Fe,X=0,005;Al,X=0,01;Cr, X=0,01), $%
Bi_{1,7}Pb_{0,2}Sb_{0,1}Sr_{2}Ca_{2}Cu_{3}O_{Y}$ and $%
Bi_{1,5}Pb_{0,3}Sb_{0,2}Sr_{2}Ca_{2}Cu_{3}O_{Y}$ ceramics. It is found that
step-like regions are revealed on AC magnetic field dependence curves $%
T_{m}^{g}(h_{0})$ of hysteresis losses peak temperature for superconducting
granuls. The steps at temperature scale depend on sample's composition as
well as on the type of atoms substituted for the copper. They range from 1
to 2 K for Y-based and about 5 K for Bi-based samples. 

Possible interpretation of the obtained results is presented.\bigskip 

\bigskip
\end{abstract}

\section{Introduction}

The dependence of critical parameters of high-temperature superconductors
(HTSCs) as a function of external influence is of a greate interest.
Particularly, the nature of this dependence is due to the existence of
structural inhomogeneities in HTSCs and the knowleage of their origine,
promotes to elucidate the mechanism of high-temperature superconductivity.

By now, there are several works, where the influence of inhomogeneities on
critical parameters of Y-based and Bi-based HTSCs was studied by using
different measurements [1-4]. In refs. \cite{1,3,4} the influence of
inhomogeneities on the ordering parameters was studied in superconducting
ceramics, films and single crystals by measuring of the electroresistivity,
above the critical temperature $T_{c}$.

It is worth to note, that the irradiation of HTSCs by high energy particles
affects the structural inhomogeneities inside the granules. It was shown in
ref. \cite{5} that for ceramic $YBa_{2}Cu_{2.99}Na_{0.01}O_{Y}$ samples,
irradiated by 8 MeV electrons with doses from $10^{14}$ to $2\times 10^{18}$ 
$e/cm^{2}$, step-like regions appeared on the initial AC magnetic field
dependence curve of intragranular hysteresis losses peak temperature $%
T_{m}^{g}$ , while in \cite{6} these regions disappeared after electron
irradiation of Y-based ceramics.

In this work we have shown the existence of structural inhomogeneities in $%
YBa_{2}Cu_{3-X}M_{X}O_{Y}$ (M=Fe,X=0,005; Al, X=0,01; Cr,X=0,01), $%
Bi_{1,7}Pb_{0,2}Sb_{0,1}$ $Sr_{2}Ca_{2}Cu_{3}O_{Y}$ and $%
Bi_{1,5}Pb_{0,3}Sb_{0,2}Sr_{2}Ca_{2}Cu_{3}O_{Y}$ compounds and that their
effect on the magnetic properties is due to sample's composition.

\section{Results and diccussion}

AC magnetic susceptibility measurements were performed by means of a
equipment, the details of which were presented in refs. \cite{7,8}. The
temperature $T$ of the samples ranges from 78 to 100 K for Y-based and from
78 to 120 K for Bi-based samples.

$YBa_{2}Cu_{3-X}M_{X}O_{Y}$ (M=Fe, X=0,005; Al, X=0,01; Cr, X=0,01), $%
Bi_{1,7}Pb_{0,2}$ $Sb_{0,1}Sr_{2}Ca_{2}Cu_{3}O_{Y}$ and $%
Bi_{1,5}Pb_{0,3}Sb_{0,2}Sr_{2}Ca_{2}Cu_{3}O_{Y}$ superconducting samples
were prepared by the conventional solid state reaction technique.
Subsequently, they will be referred to as Y-Fe, Y-Al, Y-Cr, B1 and B2,
respectively (see the table). The samples had cylindrical form. The values
of the bulk density, onset of superconducting transition temperature $%
T_{c}^{on}$ and decreasing rate $\mid dT_{m}^{g}/dh_{0}\mid $ (detected for
initial low magnetic field in the linear dependence region) for the samples
are presented in the table. The temperature of samples was detected within $%
\pm $ $0.1$ $K$ by measuring a copper wire resistance. We have determined by
examining $T_{c}^{on}$ the higher temperature inflection point of the $\chi
^{\prime }(T)$ curve for AC magnetic field amplitude $h_{0}$ $=$ $10$ $mOe$
(fig.1, curve $c$), while $T_{m}^{g}$ was obtained from the position of the
higher temperature peak of the imaginary part of magnetic susceptibility $%
\chi ^{\prime \prime }(T)$ (fig.1, curves $a$ and $b$; fig.2). We have
chosen the frequency of applied AC magnetic field to be $1$ $kHz$, with the
amplitude $h_{0}$ varying in the range from 10 mOe to 30 Oe.

$T_{m}^{g}(h_{0})$ curves for Bi-based and Y-based samples are plotted in
figs. 3 and 4, respectively. As it is seen from fig.3, there is a linearly
decreasing region on the $T_{m}^{g}(h_{0})$ curves for relatively low fields
in Bi-based samples. However, the length of this region is different
depending on the sample doping. It is known, that the value of $\mid
T_{m}^{g}/dh_{0}\mid $ characterizes the pinning force for the Abrikosov
magnetic vortices penetrating into the grains \cite{9}. The pinning force is
lower for higher values of decreasing rate $\mid T_{m}^{g}/dh_{0}\mid $.

\begin{table}[tbp] \centering%
\begin{tabular}{|c|c|c|c|c|c|}
\hline
\# & Sample & Contenet & $\frac{T_{c}^{on},\text{ K}}{(h_{0}=10mOe)}$ & $%
\frac{\rho }{g/cm^{3}}$ & $\frac{\left| dT_{m}^{g}/dh_{0}\right| }{K/Oe}$ \\ 
\hline
1 & Y-Fe & $YBa_{2}Cu_{2.995}Fe_{0.005}O_{Y}$ & 91.3 & 4.71 & 0.3 \\ \hline
2 & Y-Al & $YBa_{2}Cu_{2.99}Ale_{0.01}O_{Y}$ & 91.25 & 3.78 & 0.1 \\ \hline
3 & Y-Cr & $YBa_{2}Cu_{2.99}Ale_{0.01}O_{Y}$ & 91.85 & 3.85 & 0.25 \\ \hline
4 & B1 & $Bi_{1.7}Pb_{0.2}Sb_{0.1}Sr_{2}Ca_{2}Cu_{3}O_{Y}$ & 108.0 & 3.03 & 
0.35 \\ \hline
5 & B2 & $Bi_{1.5}Pb_{0.3}Sb_{0.2}Sr_{2}Ca_{2}Cu_{3}O_{Y}$ & 106.5 & 2.68 & 
0.2 \\ \hline
\end{tabular}
\caption{Some initial critical parameters for samples\label{key}}%
\end{table}%

In all samples, with increasing the magnetic field from some critical value $%
h_{0}$ , a transformation process of $T_{m}^{g}(h_{0})$ curve occurs, which
is accompanied by appearance of steps with different heights and widths. For
Bi-based samples this phenomenon is relatively greater than that of in
Y-based ones. Numerous investigations showed that Bi-based samples are more
sensitive to external influences than Y-based ones. The reason of such a
behaviour is that Bi-based compounds have more complicated structure and
hence contain relatively greater concentration of structural defects. It is
known that in HTSCs the CuO planes are responsible for superconductivity 
\cite{10}. The number of these planes in a unit cell of Bi-based compounds
reaches up to three, and even the lowest amount of defects induced in there
may results in s strong variations of the critical temperature. This
probably causes relatively higher steps on $T_{m}^{g}(h_{0})$ curves for B1
and B2 samples as compared with those for Y-based ones.

It should be noted that the substitution of Bi by Pb and Sb leads to the
stabilization of $Bi_{2}Sr_{2}Ca_{2}Cu_{3}O_{Y}$ $(2223)$ samples \cite{11}.
As it is seen from fig. 1,$c$ the B2 sample with higher content of Pb and Sb
at relatively low temperature has greater diamagnetism, and the
superconducting transition curve $\chi ^{\prime }(T)$ is sharpen than that
of for B1. At the same time, the peaks on the $\chi ^{\prime \prime }(T)$
curve for B2 sample become narrower too. Such a behaviour of AC
susceptibility indicates that the addition of Pb and Sb impurities increases
the homogeneity of the samples \cite{12}. This can be seen also from fig. 3
where the initial slop of $T_{m}^{g}(h_{0})$ curve decreases with increasing
the concentration of Pb- and Sb- doping. The latter means an increase of the
pinning force for the Abrikosov vortices. Such a behaviour coincides with
ref. \cite{11}, where the Pb and Sb doping leads to the increase of $2223$
phase fraction in the sample and results in the stabilization of it's
magnetic properties.

It is known that a superconducting granule has a complex structure, as well
as critical parameters close to those for single crystals \cite{12}. The
detailed investigation of intragranular structure of Y-based samples was
performed in \cite{13} by means of scanning electron microscopy. These
results suggested that the granules have polyedric forms with an average
size of a several microns. The porosity is the main feature of granules.
Besides the structural defects, as well as the nonsuperconducting $%
Y_{2}BaCuO_{5}$ phase and the majority of amorphous phase are mainly
concentrated in subsurface layer of granule. It was suggested in \cite{14}
that during the sintering of Bi-based samples firstly a low temperature $%
Bi_{2}Sr_{2}CaCu_{2}O_{Y}$ $(2212)$ phase is formed, which is further
wrapped by a layer of $2223$ with higher critical temperature. In B1 sample
the slop of $T_{m}^{g}(h_{0})$ for $h_{0}$ up to 15 Oe is due to the pinning
force of the Abrikosov vortices penetrating into the $2223$ layer, while for
higher values of $h_{0}$ this slope may be attributed to the motion of
vortices through the layer consisting of $2212$ phase inside the granule.
The latter phase is more sensitive to the application of a magnetic fields
and an inflection point on $T_{m}^{g}(h_{0})$ curve is observed near the $%
h_{0}$ $=$ $15$ $Oe$ . As the Pb- and Sb- doping increases the $2223$ phase
fraction, the B2 sample becomes more stable in respect to a magnetic field
which is partially manifested in a shift of the corresponding inflection
point on the $T_{m}^{g}(h_{0})$ curve from 15 to the 25 Oe. Besides, for B2
sample higher $T_{m}^{g}$ values are observed in magnetic field $17$ $\leq $ 
$h_{0}$ $\leq $ $28$ $Oe$ region in compared with the B1 sample. The
structure of the subsurface layer of granule, indeed, varies depending on
sample's content. This leads to the formation of layers around the granule
with different values of critical parameters, which cause the observed
step-like regions on $T_{m}^{g}(h_{0})$ curves.

Fig. 4 shows that for Y-Fe sample the number and forms of steps on $%
T_{m}^{g}(h_{0})$ curve differ from those for other samples. This difference
may be due to the relatively higher bulk density of Y-Fe sample (see the
table), which results in the oxygen deficiency of the compound. The latter
make the influence of microinhomogeneities caused by the Fe doping on the
sample's superconducting properties more expressed \cite{15}. This
confirmation is supported by fig. 2, which shows, that for Y-Fe sample
(curve $a$) the inhomogeneity is the highest, because it's $\chi ^{\prime
\prime }(T)$ peak is the most broaden \cite{12}. It is not ruled out that
the above mentioned difference might be also due to the induced several
oxygen environments of Fe, substituted for Cu atoms (see ref. [8] and
references therein). 

It was shown \cite{6}, that the observed minima on the initial $%
T_{m}^{g}(h_{0})$ curves for $YBa_{2}Cu_{3}O_{Y}$ , $%
YBa_{2}Cu_{2.99}Fe_{0.01}O_{Y}$ and $YBa_{2}Cu_{2.99}Ni_{0.01}O_{Y}$ samples
disappeared after irradiation by 8 MeV electrons with dose $2\times 10^{16}$ 
$e/cm^{2}$. One of the reasons of the observed phenomenon is that the
electron irradiation can redistribute the initial structural defects in
sample and create new ones, leading to the variation of crystalline lattice
ordering. This is consistent with the result that the electron irradiation
induces the uniform distribution of structural defects in a sample \cite{10}.

\section{Conclusions}

From the above mentioned we can conclude that the appearance of the steps on
the $T_{m}^{g}(h_{0})$ curves may be attributed to the existence of
different structural defects in granules, which can act as a flux pinning
centers with definite pinning force values. These centers are responsible
for the penetration of the Abrikosov vortices into the granules. The
existence of ''weak'' and ''strong'' pinning centers is due to inhomogeneous
distribution of the structural defects, and their coexistence leads to the
inhomogeneous penetration of magnetic fields into the granule. This causes
the formation of the step-like regions on $T_{m}^{g}(h_{0})$ curves. The
effect is better expressed in Bi-based samples than in Y-based ones. This
difference may be explained by relatively complex structure of Bi-based
samples. The step-like behaviour of $T_{m}^{g}(h_{0})$ curve for the Y-Fe
sample is distinctly expressed among the Y-based samples. Such a behaviour
may be due to the relatively higher content of defects, created by the
oxygen deficiency because of the higher bulk density of the Y-Fe sample.

Further structural measurements are needed for exact interpretations of
obtained results.

\end{document}